\documentclass{elsarticle}

\usepackage[english]{babel}           
\usepackage{amsmath,amssymb,amsfonts} 
\usepackage{graphicx}                 
\usepackage[utf8]{inputenc} 
\usepackage[T1]{fontenc}    

\providecommand{\floor}[1]{\left \lfloor #1 \right \rfloor}
\newcommand{\algebra}[1]{\ensuremath{\mathcal{#1}}}
\newcommand{\Sphere}{\ensuremath{\mathbb{S}^2} }
\newcommand{\FuzzySphere}{\ensuremath{\mathbb{S}_{\text{F}}^2}}
\makeatletter
\newcommand{\lambdabar}{{\mathchoice
  {\smash@bar\textfont\displaystyle{0.25}{1.2}\lambda}
  {\smash@bar\textfont\textstyle{0.25}{1.2}\lambda}
  {\smash@bar\scriptfont\scriptstyle{0.25}{1.2}\lambda}
  {\smash@bar\scriptscriptfont\scriptscriptstyle{0.25}{1.2}\lambda}
}}
\newcommand{\smash@bar}[4]{%
  \smash{\rlap{\raisebox{-#3\fontdimen5#10}{$\m@th#2\mkern#4mu\mathchar'26$}}}%
}
\makeatother
\newcommand{\inner}[2]{\ensuremath{\left\langle #1, #2\right\rangle}}
\newcommand{\textsubscript}[1]{\ensuremath{_{\textnormal{#1}}}}
\newcommand{\manifold}{\ensuremath{\mathcal{M}}}
\newcommand{\boundary}{\partial}
\newcommand{\fmanifold}{\ensuremath{\mathcal{M}_\text{F}}}
\DeclareMathOperator{\dimension}{dim}
\DeclareMathOperator{\volume}{vol}

\DeclareMathOperator{\Tr}{Tr}
\newcommand{\reals}{\ensuremath{\mathbb{R}}}
\newcommand{\naturals}{\ensuremath{\mathbb{N}}}
\newcommand{\cmplxset}{\ensuremath{\mathbb{C}}}
\newcommand{\imag}{\textrm{i}}
\newcommand{\dif}{\textrm{d}}
\newcommand{\unit}[1]{\,\text{#1}}

\title{A note on the fuzzy sphere area spectrum, black hole
  luminosity, and the quantum nature of spacetime}
  
\author[psu,ufc]{Victor~Santos}
\ead{vsantos@gravity.psu.edu}
\author[ifpb]{C.~A.~S.~Silva}
\ead{calex@fisica.ufc.br}
\author[ufc]{C.~A.~S.~Almeida}
\ead{carlos@fisica.ufc.br}

\address[psu]{Institute for Gravitation and the Cosmos \& Physics Department\\%
The Pennsylvania State University, University Park PA 16802, U.S.A.}
\address[ufc]{Departamento de F\'isica -- Universidade Federal do Cear\'{a},\\%
 C.P.~6030, 60455--760, Fortaleza-CE, Brasil}
\address[ifpb]{Instituto Federal de Educa\c{c}\~ao, Ci\^encia e Tecnologia da Para\'iba(IFPB),\\%
Campus Campina Grande -- Rua Tranquilino Coelho Lemos, 671, Jardim Dinam\'erica I.}

\begin{document}
\begin{abstract}
  Non-commutative corrections to the classical expression for the
  fuzzy sphere area are found out through the asymptotic expansion for
  its heat kernel trace. As an important consequence, some quantum
  gravity deviations in the luminosity of black holes must appear. We
  calculate these deviations for a static, spherically symmetric,
  black-hole with a horizon modeled by a fuzzy sphere. The results obtained could be
  verified through the radiation of black holes formed in the Large
  Hadron Collider~(LHC).
\end{abstract}
\maketitle

\section{Introduction} 
\label{sec:introduction}

The search for the understanding of the quantum nature of spacetime is one of the most
difficult challenges Physics has faced. Several theories have been
applied for reach this mission.  Even though we do not have yet a
complete theory of quantum gravity, some results obtained by
approaches like String theory and Loop Quantum Gravity can give us
some clues about the spacetime behavior in the Planck regime.  Among
the most important results, we have that the quantum nature of
spacetime is somehow related with a noncommutative behavior of the
geometry in the Planck scale.

In fact, the use of non-commutative spaces in order to investigate the
quantum behavior of physical systems is not new in Physics, being the
quantum mechanical phase space a prime example. Actually, Heisenberg
was the first to suggest extending noncommutativity to the coordinates
of the physical space as a possible way of removing the infinite
quantities appearing in field theories in his letters to
Ehrenfest~\cite{1985-Heisenberg--382} and
Peierls~\cite{1979-Pauli--}. At that time Heisenberg could not
formulate this idea mathematically. The first papers on the subject
was published in 1947 by Snyder~\cite{1947-Snyder-PR-38,
  1947-Snyder-PR-68} and Yang~\cite{1947-Yang-PR-874}, regaining
attention over the last decades~\cite{1995-Connes--, 1999-Madore--,
  1998-Connes-JHEP-3, 2000-Castellani-CQG-3377} due to its appearance
in some prominent quantum gravity frameworks, like String
theory~\cite{1999-Seiberg-JHEP-32, 1998-Cheung-NPB-185,
  1999-Chu-NPB-151, 1999-Schomerus-JHEP-30, 2000-Seiberg-JHEP-21,
  2000-Seiberg-JHEP-44} and Loop Quantum
Gravity~\cite{2011-Baratin-CQG-175011, 2011-Aastrup-CQG-75014,
  2010-Denicola-CQG-205025}.

The motivation behind such spaces is that they have a potential for
replacing the classical geometric description of spacetime by
incorporating quantum fluctuations as follows: a modified commutation
relation for the spacetime coordinates,
\begin{equation}
[x^{\mu}, x^{\nu}] = \imag\theta^{\mu\nu},
\end{equation}
incorporates a spacetime uncertainty
relation~\cite{1929-Robertson-PR-163}
\begin{equation}
\Delta x^{\mu}\cdot\Delta x^{\nu} \geq \frac{1}{2}|\theta^{\mu\nu}|,
\end{equation}
which spoils the intuitive notion of point at distances shorter
than~$\sqrt{\theta}$. This is suspected to happen if we probe
distances of the order of the Planck's length~$l_{\text{P}}\sim
10^{-35}\unit{m}$, when gravitational quantum effects are not
negligible. Besides, the semiclassical limit of quantum gravity
theories is expected to be described by a field theory on some
non-commutative spacetime~\cite{2006-Freidel-PRL-221301}.

Among the realizations of non-commutative geometry there are the
so-called \emph{fuzzy spaces}. A fuzzy space can be roughly defined as
a quantum representation of a manifold. Its construction is based on
the fact that coadjoint orbits of Lie groups are symplectic manifolds,
and hence can be quantized~(under certain conditions). The
quantization procedure introduces a parameter analogue to the Planck's
constant, being the classical manifold limit recovered when this
parameter goes to zero. In the case where the Lie group is compact and
semi-simple, its quantum version is compact and identified as a
finite-dimensional matrix algebra on which the group acts. Moreover,
it preserves the symmetries of its classical version. Then a fuzzy
space, realized as a matrix algebra, provides a concrete method to
model the spacetime noncommutativity.


A possible way to verify the noncommutative properties of the
spacetime is to investigate quantum gravity corrections nearby a black
hole. By the way, the
use of noncommutative geometry to address black hole physics is mainly
motivated by the hope that black holes might play a major role in our
attempts to shed some light on the quantum nature of gravity such as
the role played by atoms in the early development of quantum
mechanics. In this way, noncommutativity has been used to investigate
possible quantum gravity effects in black hole physics too \cite{Nicolini:2008aj}.
It includes some fuzzy spaces approaches, where fuzzy manifolds have
been successfully used as a way to quantize the event horizon of a
black hole~\cite{2005-Dolan-JHEP-8,
  2009-Silva-PLB-318,2011-Silva-EPL-10007}. In the last two 
references, using the Hopf algebra structure of the noncommutative
manifold to model a topology change in black-hole, the authors argued
that a solution for the information loss paradox can be obtained. 


Due to the possibility of the formation of micro black holes in the
Large Hadron Collider~(LHC), it is interesting to investigate how
quantum gravity effects could appear in the radiation emitted by these
black holes.  Following the idea that noncommutative geometry can be
used to model these effects, in this work, we shall address how fuzzy
spaces can be used to investigate quantum gravity corrections in the
black hole luminosity, modeling the black hole event horizon as a
fuzzy sphere.

A crucial point is to know the quantum geometrical features of the
manifold which has been used to model the black hole event
horizon. In this way, several approaches in order to investigate the
correct spectrum of the fuzzy sphere geometrical quantities have been
proposed \cite{2009-Silva-PLB-318, 2011-Silva-EPL-10007, Hod:1998vk, Banerjee:2010be, Rovelli:1994ge}. 
To address how the quantum gravity fluctuations in
the black hole horizon will modify the black hole luminosity, the
method we will use in the present work will consist in to investigate the fuzzy
sphere area spectrum through the asymptotic expansion for the heat
kernel trace.

The major motivation to use this method is related to the fact that,
since a fuzzy space is defined in terms of the spectral triple, its
geometrical properties must be recovered algebraically. Actually,
following the standard procedure for the classical sphere, its
geometry can be determined by the knowledge of the Laplacian spectrum.
In this way, our first goal here is to extrapolate such approach to a
fuzzy space, in order to investigate its effects on the classical area
formula. As we will see in the following sections, such extrapolation
leads to an exponential correction to the sphere area, which vanishes
either in the commutative limit or in low-energy limit.

This paper is organized as follows. Section~\ref{sec:prelim} is
devoted to organize the main concepts used in our discussions, mainly
the relation between the heat kernel trace and the geometrical
area. Section~\ref{sec:area_spectrum} contains the actual area
spectrum derivation and, in Section~\ref{sec:bh_luminosity}, we
employ the derived area spectrum to discuss modifications in the
luminosity of a Schwarzschild black-hole. After this, we make the
final remarks.

For the physical quantities, we employ dimensionless units such that
\begin{equation}
\hslash = c = G = k\textsubscript{B}= 1.
\end{equation}
With these definitions all quantities are written in terms of the Planck
mass~$m_{\text{P}} = 2.176\times 10^{-8}\unit{kg}$,
length~$l_{\text{P}} = 1.616\times 10^{-35}\unit{m}$,
time~$t_{\text{P}} = 5.381\times 10^{-44}\unit{s}$, 
energy~$E_{\text{P}} = 1.956\times 10^{9}\unit{J}$, etc. As an example, the
luminosity of the sun can be written
as~$L_{\odot}=1.1\times 10^{-26}t_{\text{P}}/E_{\text{P}}$.

\section{Preliminaries} 
\label{sec:prelim}
In the present section we review some concepts of the classical geometry of
Riemannian manifolds which will help us to calculate the
non-commutative corrected area of a fuzzy sphere. Throughout this
section, let~$\manifold$ denote a smooth (orientable) Riemannian
manifold with metric~$g$ of dimension~$n=\dimension(\manifold)$.

%
%
\subsection{Heat kernel} 
\label{subsec:prelim.heat_equation}
By definition, the heat equation on~\manifold\ is defined as
\begin{equation}
\frac{\partial u}{\partial t} + \Delta_2 u = 0,\quad u:[0,\infty) \times 
\manifold\rightarrow\reals,
\label{eq:heat_equation_manifold}
\end{equation}
where~$\Delta_2$ is the Laplace-Beltrami operator with respect to the
second variable (the spatial variable over the
manifold~\manifold). The function~$u(t,x)$ can represent the
temperature at the point~$x\in\manifold$ at time~$t$.

Given an initial distribution~$u(t=0,x)=f(x)$ for~$x\in\manifold$ and
a Dirichlet condition at the boundary~$\boundary\manifold$
of~$\manifold$,
\begin{equation}
u(t,x) = 0,\text{ if } x \in \boundary\manifold,
\label{eq:dirichlet_condition_heat_equation}
\end{equation}
there is a fundamental solution for~\eqref{eq:heat_equation_manifold},
called the \emph{heat kernel}, which is a function
\begin{equation}
K:(0,\infty) \times \manifold \times \manifold \longrightarrow \manifold
\label{eq:heat_kernel_definition}
\end{equation}
satisfying the following conditions
\begin{enumerate}
\item $K(t,x,y)$ is $C^1$ in $t$ and $C^2$ in $x$ and $y$;
\item $K$ solves the heat equation \eqref{eq:heat_equation_manifold}:
\begin{equation}
\frac{\partial K}{\partial t} + \Delta_2 K = 0;
\end{equation}
\item $K$ satisfy the boundary condition $K(t,x,y) = 0 \iff x \in
\boundary\manifold$;
\item the equality 
\begin{equation}
\lim_{t\to 0^+}\int_\manifold \volume_y(\manifold)\, K(t,x,y)f(y) = f(x),
\end{equation}
where the subscript in~$\volume_y$ denotes the variable integration,
holds uniformly for every function~$f$ continuous on~$\manifold$ and
vanishing on~$\boundary\manifold$.
\end{enumerate}

\subsection{Heat trace} 
\label{subsec:prelim.heat_kernel_trace}

Once we have defined properly the Laplace operator~$\Delta$
on~\manifold, its spectrum~$\{\lambda_{\alpha}\}_{\alpha\in I}$ for an
arbitrary index set~$I$ together with its eigenfunctions~$f_{\alpha}$,
\begin{equation} 
\Delta f_{\alpha} = \lambda_{\alpha} f_{\alpha},
\label{eq:laplacian_eigenvalue_equation}
\end{equation}
determines the unique heat kernel~\eqref{eq:heat_kernel_definition} as follows:
\begin{equation}
K(t,x,y) = \sum_{\alpha\in I}e^{-\lambda_\alpha t}f_{\alpha}(x)f_{\alpha}(y),
\label{eq:sl_decomposition}
\end{equation}
with absolute and uniform convergence for~$t > 0$.

The \emph{heat trace} of the heat kernel is defined by
\begin{equation}
\theta(t)=\int_M\volume_x(M)\,K(t,x,x),
\label{eq:heat_trace_definition}
\end{equation}
which can be written in a more simple form using
decomposition~\eqref{eq:sl_decomposition}:
\begin{equation}
\theta(t)=\sum_{\alpha\in I}e^{-\lambda_\alpha t},
\label{eq:heat_trace_redefinition}
\end{equation}
where the summation is made over all eigenvalues, taking into account
its multiplicity.

\subsubsection{Asymptotic expansion and geometrical quantities} 
\label{subsubsec:prelim.heat_kernel_trace.asymptotic_expansion}

The heat trace~\eqref{eq:heat_trace_redefinition} has an asymptotic
expansion in the region~$t\to 0^+$, as shown
in~\cite{1994-Gilkey--}:
\begin{equation}
\theta(t)=(4\pi t)^{-n/2} \sum_{i=0}^{N}c_i t^{i/2}%
+\mathcal{O}\big(t^{(N+1)/2}\big),
\label{eq:heat_trace_asymptotic_expansion}
\end{equation}
where~$\mathcal{O}$ denotes the condition
\begin{equation}
f(t)=\mathcal{O}(g(t))  \iff \exists k\in\reals : %
\bigg|\frac{f(t)}{g(t)}\bigg| < k \text{ when }t \to 0,
\label{eq:big_O_notation}
\end{equation}
and the coefficients~$c_i$ are called Seeley-de~Witt coefficients, and
depends on the geometry of the manifold $\manifold$. If~$\manifold$ is
compact and has a compact $(n-1)$-dimensional
boundary~$\mathcal{B}=\boundary\manifold$, the first few coefficients
of~\eqref{eq:heat_trace_asymptotic_expansion} are explicitly shown to
encode the following geometric
information~\cite{1967-Mckean-JoDG-43}:
\begin{equation}
c_0 = \text{vol}(\manifold),\quad 
c_1 = -\frac{\sqrt{\pi}}{2}\text{vol}(\mathcal{B}),\quad
c_2 = \frac{1}{3}\int_\manifold S - \frac{1}{6}\int_\mathcal{B} J,
\label{eq:expansion_coefficients}
\end{equation}
where~$S$ is the scalar curvature of~$\manifold$ and~$J$ is the mean
curvature at its boundary~$\boundary\manifold$.

\subsection{The fuzzy sphere} 
\label{subsec:prelim.fuzzy_sphere}

The triple~$(\mathcal{H},\mathcal{A},\Delta)$
where~$\mathcal{H}=L^2(\manifold)$ is the Hilbert space of
square-integrable functions on~$\manifold$, $\Delta$ is the
Laplace-Beltrami operator
and~$\mathcal{A}=\mathcal{C}^\infty(\manifold)$ is the algebra of
smooth (bounded) functions on~$\manifold$, was shown to contain all
geometrical information about the manifold; that is, the geometry
of~$\manifold$ can be formulated in terms
of~$\mathcal{A}$~\cite{1994-Connes--}.

In a similar way, the quantum (fuzzy) version~$\fmanifold$
of~$\manifold$ can be defined by a sequence of triples
\begin{equation}
\fmanifold := \big\{(\mathcal{H}_N,\mathcal{A}_N,\Delta_N)\big\}_{N\in\naturals}
\label{eq:fuzzy_triple_sequence}
\end{equation}
parametrized by a natural number~$N$,
where~$\mathcal{A}_N=\text{Mat}_{N}({\cmplxset})$ is the algebra of
$N\times N$ matrices with complex entries,
$\mathcal{H}_N=\cmplxset^{N^2}$ is the Hilbert space over
which~$\mathcal{A}_N$ acts and~$\Delta_N$ is a representation of the
Laplace operator on the matrix space. The Hilbert space inner product
is
\begin{equation}
\inner{\Phi}{\Psi} = \frac{1}{N}\Tr(\Phi^\dagger\,\Psi).
\end{equation}

The fuzzy sphere $\FuzzySphere$~\cite{1992-Madore-CQG-69} is a very
simple example of a fuzzy space, which can appear as vacuum solutions
in Euclidean gravity~\cite{2006-Sheikh-Jabbari-PLB-119,
  2003-Abe-PRD-25002}. A simple recipe to obtain~$\algebra{A}_N$ is to
replace the global coordinates~$x_1$, $x_2$, $x_3$ of the
sphere~$\Sphere$ of radius~$R$, which satisfy
\begin{equation}
x_1^2 + x_2^2 + x_3^2 = R^2,
\end{equation}
by a set of operators~$X_1$, $X_2$, $X_3$ satisfying
\begin{equation}
\sum_{a=1}^{3} (X_a)^2 = R^2\,\text{Id}
\end{equation}
(where~$\text{Id}$ is the identity operator), whose representation is
proportional to the unitary irreducible
representation~$j\in\naturals/2$ of the rotation generators~$J_a$ of the $SU(2)$ algebra
\begin{equation}
[J_a, J_b] = \imag\,\epsilon_{abc}\,J_c,\quad
\sum_{a=1}^3(J_a)^2=j(j+1)\,\text{Id}.
\end{equation}
From this we can see that~$X_a$ are represented by~$(2j+1)\times(2j+1)$
hermitian matrices satisfying
\begin{equation}
X_a=\lambdabar J_a,\quad 
[X_a, X_b]=\imag\lambdabar\,\epsilon_{abc}X_c,
\label{eq:coord_commut}
\end{equation}
with
\begin{equation}
\lambdabar=\frac{R}{\sqrt{j(j+1)}},
\label{eq:quantization_parameter}
\end{equation}
and that its components can be decomposed into~$2j+1$ irreducible
representations of $SU(2)$ algebra with angular
momentum~$\ell=0,1,\dots,2j$. Thus, defining the index
in~\eqref{eq:fuzzy_triple_sequence} as~$N=2j+1$,
equation ~\eqref{eq:quantization_parameter} is rewritten as
\begin{equation}
\lambdabar=\frac{R}{\sqrt{N^{2}-1}}.
\end{equation}
The parameter~$\lambdabar$ has dimension of length, and plays a role
analogue to the Planck's constant in quantum mechanics, as a
quantization parameter. From~\eqref{eq:coord_commut} we can see that
in the limit~$\lambdabar\to 0$ ($N\to\infty$) the matrices~$X_a$
become commutative, and we recover the standard commutative sphere.

\subsubsection{Laplace operator}
\label{subsubsec:prelim.fuzzy_sphere.laplacian}

The Laplace operator~$\Delta_N$ can be defined over the fuzzy sphere
in terms of the adjoint action of the
generators~$J_a$~\cite{2003-OConnor-MPLA-2423}:
\begin{equation} 
\Delta_N\Psi=\frac{1}{R^2}\sum_{a=1}^{3}
\Big[J_a^{(N)},\Big[J_a^{(N)},\Psi\Big]\Big];
\label{eq:fuzzy_laplacian}
\end{equation}
then, solving the eigenvalue
equation~\eqref{eq:laplacian_eigenvalue_equation} we can find the same
spectrum of the usual commutative Laplacian, except by a cut-off
at~$\ell=N-1$; that is,
\begin{equation} \label{eq:nc_laplacian_spectrum}
\lambda_{\ell} = \frac{\ell(\ell+1)}{R^2},\quad
\ell=0,1,\cdots,N-1.
\end{equation}
This spectrum coincides up to order~$N-1$ with the usual continuum
counterpart, when the Laplace operator acts on the space of functions
on a sphere.

\section{Fuzzy Sphere Area Spectrum} 
\label{sec:area_spectrum}

In this section, we show how the heat kernel trace can be used to
evaluate the area of a fuzzy sphere, extrapolating the classical
results in the quantum regime.  Such extrapolation will be used as a
\emph{physical} artifact to find signatures of the noncommutativity
in the classical realm, revealing some features of the spacetime in
the quantum regime which must be described by a full quantum gravity
theory.

To begin with, we have that, since the Laplace operator spectrum is
given by~\eqref{eq:nc_laplacian_spectrum}, the heat
trace~\eqref{eq:heat_trace_redefinition} for the fuzzy sphere can be
written as the finite sum
\begin{equation}
\theta(t)=\sum_{0\leq \ell<N}(2\ell+1)e^{-\lambda_{\ell}t},
\label{eq:fuzzy_sphere_heat_trace}
\end{equation}
where the factor~$(2\ell+1)$ takes into account the multiplicity of each
eigenvalue $\lambda_{\ell}$.

Since~$\ell$ is a pure number, the
eigenvalues~\eqref{eq:nc_laplacian_spectrum} have dimension
of~[length]$^{-2}$, and hence the expansion parameter~$t$ has
dimension of~[energy]$^{-2}$ in natural units; defining
\begin{equation}
\Lambda = \frac{1}{\sqrt{t}} 
\label{eq:energy_scale}
\end{equation}
we have that~$\Lambda$ has dimension of energy. We interpret such
parameter as the energy scale where we probe the fuzziness of the
geometry.

With such interpretation, comparing the heat trace with the spectral
action of Connes', non-commutative geometry leads to a reasonable
cutoff for~$\Lambda$, of the order of the Planck scale:
\begin{equation}
\Lambda = \Lambda_\ast \sim M\textsubscript{P}\;. 
\end{equation}
Thus, conversely of the commutative case, the asymptotic
expansion~\eqref{eq:heat_trace_asymptotic_expansion} must be taken not
at~$t\to 0$ but at a finite value~$t\to
t_{\ast}=M\textsubscript{P}^{-2}$. Hence, if we define
\begin{equation} 
\alpha(s) = 4\pi t\,\theta(t) = 4\pi t\sum_{0\leq k<N}
(2k+1)e^{-\lambda_{k}t}.
\label{eq:regularized_heat_trace}
\end{equation}
The extrapolation of the classical behavior of the heat trace shown in
equations~\eqref{eq:heat_trace_asymptotic_expansion}
and~\eqref{eq:expansion_coefficients} relates the area~$A$ to the
coefficient of the power~$(t-t_{\ast})^{0}$ of the power series
expansion of~\eqref{eq:regularized_heat_trace}.

In order to evaluate such coefficient, we employ the Euler-Mclaurin
summation formula
\begin{equation} \label{eq:euler_summation_formula}
\sum_{a\leq k<b}f(k)=\int_{a}^{b}\dif x\,f(x)+\sum_{1\leq k\leq M}
\frac{b_{k}}{k!} f^{(k-1)}(x)\Big|_{a}^{b} + R_{M},
\end{equation}
where~$a$ and~$b$ are integers such that~$a\leq b$, $b_k$ denote the
Bernoulli numbers
\begin{equation} \label{eq:bernoulli_numbers}
  b_0 = 1,\ b_1 = -\frac{1}{2},\ b_2 = \frac{1}{6},\ \dots
\end{equation}
and~$R_{M}$ denotes the remainder when we truncate the series at
order~$M\geq 1$:
\begin{equation} \label{eq:remainder_euler_summation}
  R_{M} = \frac{(-1)^{M+1}}{M!}\int_{a}^{b}\dif x\,B_{M}(\{x\})f^{(M)}(x),
\end{equation}
where~$B_M(\{x\})$ is the Bernoulli polynomial and~$\{x\}$ is a
shorthand for the fractional part
\begin{equation} \label{eq:fractional_part}
\{x\}=x-\floor{x}\ (\floor{x}\text{ is the largest integer not greater than } x).
\end{equation}
Then, taking
\begin{equation}
f(x) = 4\pi t\,(2x+1)\exp\Big(-\frac{x(x+1)}{R^2} t\Big),
\end{equation}
we can verify that due to the factor~$4\pi t$, all
derivatives~$f^{(i)}(x)$ will contribute to terms involving positive
powers of~$t$, and hence the only term which contributes to the
power~$t^{0}$ of the Taylor expansion is the integral on the
right-hand side of~\eqref{eq:euler_summation_formula}; thus, the area
spectrum~$A^{\text{F}}_N$ for the fuzzy sphere is simply the first
term of~\eqref{eq:euler_summation_formula}
\begin{equation}
A^{\text{F}}_N = %
4\pi R^{2}\bigg(
1 - e^{-\frac{N(N+1)}{\Lambda_{\ast}^{2}R^{2}}}\bigg).
\label{eq:fuzzy_sphere_area_spectrum}
\end{equation}

\begin{figure}
\centering
\includegraphics[scale=0.5]{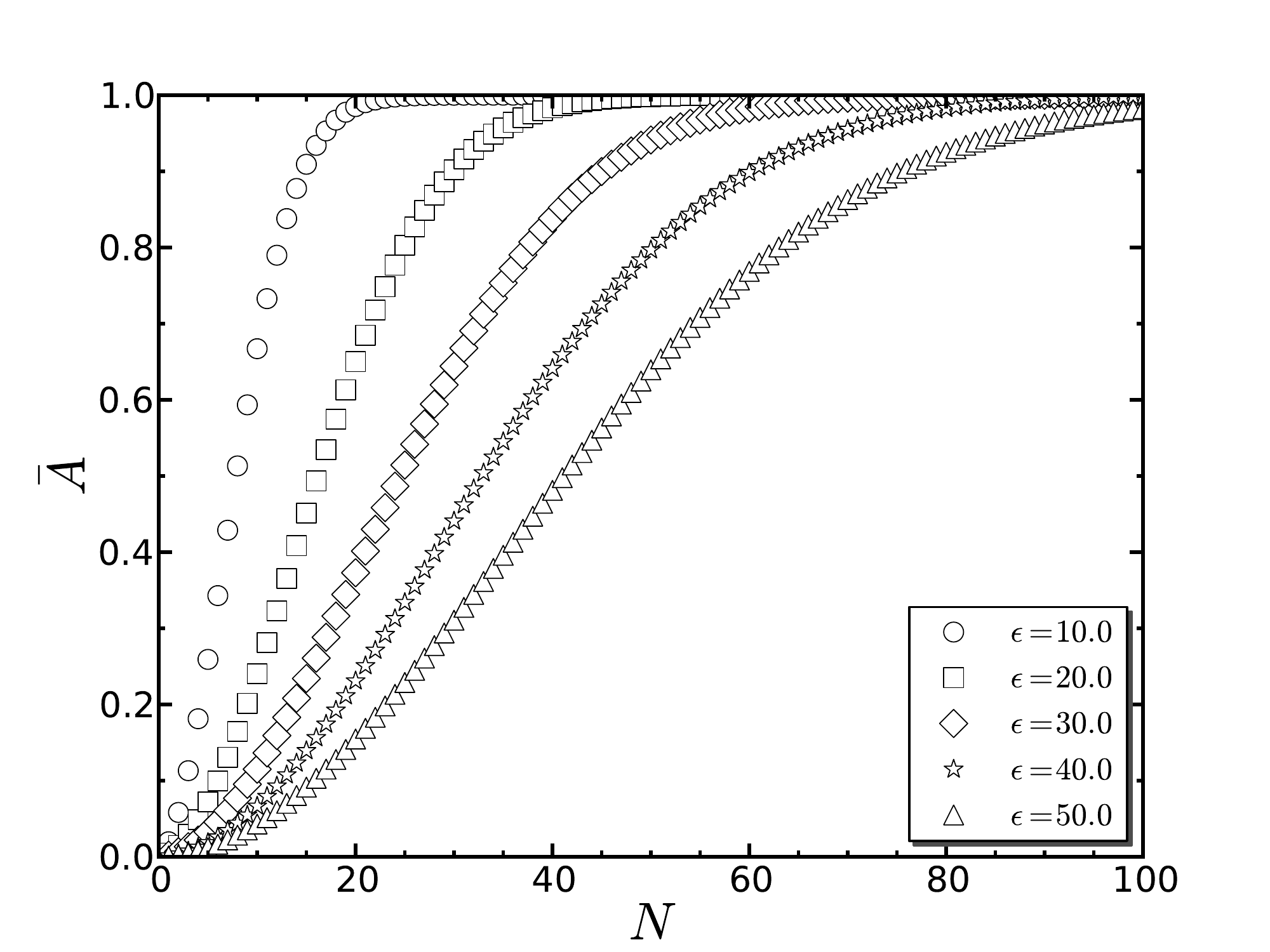}
\caption{Normalized area spectrum~\eqref{eq:normalized_area_spectrum}
  as a function of~$N$, for a fixed energy scale cutoff
  scale~$\epsilon$. Notice the effect of the cutoff in the
  significance of the noncommutativity on the area spectrum, where
  the quantum corrections contribute increasingly.}
\label{fig:norm_area_spec}
\end{figure}

To give us a better insight of this formula, we define a normalized
area spectrum, by scaling~\eqref{eq:fuzzy_sphere_area_spectrum} with
the classical area:
\begin{equation}
\bar{A}_{N} = \frac{A^{\text{F}}_{N}}{4\pi R^{2}} = 1 - e^{-\frac{N(N+1)}{\epsilon^{2}}},\quad
\epsilon = \Lambda_{\ast} R
\label{eq:normalized_area_spectrum}
\end{equation}
The behavior for the spectrum~\eqref{eq:normalized_area_spectrum} is
shown in figure~\ref{fig:norm_area_spec}.
From the equation ~\eqref{eq:fuzzy_sphere_area_spectrum} we found out that the
exponential correction quickly vanishes in the commutative
limit~$N\rightarrow\infty$, and in the limit~$\Lambda_\ast\rightarrow
0$, where the fundamental scale goes to zero, agreeing with the
classical expansion~$t\rightarrow 0$ for the heat trace.


It must be emphasized the magnitude order of the noncommutativity
parameter~$N$; being~$M\textsubscript{P}\sim 1.22\times
10^{19}\unit{GeV}$, in order to observe a correction for the area of
order of~$1\%$ we should have
\begin{equation} 
\frac{N(N+1)}{R^{2}}\sim 6.87\times 10^{56}\,\text{eV}^2,
\end{equation} 
such that, using the estimated Schwarschild radius~$R\sim
10^{20}\unit{eV}^{-1}$ for the M87 black
hole~\cite{2012-Doeleman-S-355}, we conclude that~$N=10^{50}$ still
corresponds to a strong quantum regime. This partly agrees with the
interpretation for~$N$ as the number of quanta of area~$N\sim
A/\ell\textsubscript{P}$ in entropic
gravity~\cite{2013-Gregory-PRD-64030}.

\section{Effects on the black hole's thermodynamics and luminosity}
\label{sec:bh_luminosity}
The quantum nature of spacetime can be revealed by a black hole.
In this way, some authors have argued that the black hole event horizon can be modeled by a 
noncommutative manifold 
like a fuzzy sphere \cite{Nicolini:2008aj, 2005-Dolan-JHEP-8, 2009-Silva-PLB-318,2011-Silva-EPL-10007}.
If this is true, some modifications in the description of the black hole evaporation process 
must appear as a consequence of the fuzzy sphere spectrum \eqref{eq:fuzzy_sphere_area_spectrum}.

In fact, after a quick calculation, one can find that the temperature of a Schwarzschild black hole whose
event horizon area is given by \eqref{eq:fuzzy_sphere_area_spectrum} possess a temperature

\begin{equation}
T^{F}_{N} = \frac{T_{BH}}{\sqrt{1- e^{-\frac{N(N+1)}{\epsilon^{2}}}}}
\end{equation}

\noindent and an entropy given by

\begin{equation}
S^{F}_{N} = S_{BH}(1- e^{-\frac{N(N+1)}{\epsilon^{2}}})
\end{equation}

\vspace{5mm}
\noindent where $T_{BH}$ and $S_{BH}$ are the classical Bekenstein-Hawking temperature and entropy, respectively.

It would very becoming, given the possible formation of micro black holes in the LHC, 
to investigate the consequences that the area spectrum \eqref{eq:fuzzy_sphere_area_spectrum}
can bring to the black hole luminosity. 
We have that Page's semiclassical results points to a~$1/M^{2}$
dependence for black hole luminosity~\cite{1976-Page-PRD-198,
  1976-Page-PRD-3260, 1977-Page-PRD-2402}. These results are also
found in the equally spaced spectrum proposed by
Bekenstein~\cite{1997-Makela-PLB-115}. However we can see in
equation ~\eqref{eq:fuzzy_sphere_area_spectrum} that quantum effects can
change such behavior.



In order to investigate black hole luminosity, we will use the same method due to M\"{a}kel\"{a}
~\cite{1997-Makela-PLB-115}.
In this way, we will assume that transitions
where~$\delta\omega$ is small are preferred.
It is due to the fact that, black hole radiation involves creation of
virtual particle-antiparticle pairs near the black-hole horizon, and
'swallowing' one member of some of the pairs. The greater the
energy~$\delta\omega$ of a particle of any virtual pair is, the
smaller is the probability that the pair will live long enough that
the black hole will be able to swallow one of the virtual particles.

In this way, in terms of the minimum
energy~$\delta\omega_{0}^{\textrm{F}}$, black hole luminosity will be
given by
\begin{equation}
L^{\textrm{F}} \sim \frac{\delta \omega_{0}^{\textrm{F}}}{\tau^{\textrm{F}}} = (\delta\omega_{0}^{\textrm{F}})^{2}\;.
\end{equation}
where we have used the uncertainty relation
\begin{equation}
\tau^{\textrm{F}} \sim \frac{1}{\delta\omega_{0}^{\textrm{F}}}
\end{equation}
and the superscript 'F' indicates that the quantities are fuzzy, or
quantum corrected.
To find out the value of the energy~$\delta\omega_{0}^{\textrm{F}}$ of
a particle emitted by a Schwarzschild black hole, we have
\begin{equation}
A\textsubscript{BH}^{\textrm{F}} = 16\pi(M^{\textrm{F}})^{2};
\end{equation}
in a way that,
\begin{equation}
\delta M^{\textrm{F}} = \frac{\delta A^{\textrm{F}}}{32\pi M^{\textrm{F}}} 
\end{equation}
and
\begin{equation}
\delta\omega_{0}^{\textrm{F}} = \frac{(A_{N} - A_{N-1})}{8 \sqrt{\pi A_{N}}}.
\end{equation}
Hence, black hole luminosity will be given by
\begin{equation}
L^{\textrm{F}} \sim (\delta\omega_{0}^{2})^{\textrm{F}} = \frac{(A_{N} - A_{N-1})^{2}}{64 \pi A_{N}},
\end{equation}
and we define a normalized luminosity in an analogous way as the area:
\begin{equation}
\bar{L} = \frac{L}{4\pi R^{2}} = \frac{(\bar{A}_{N} - \bar{A}_{N-1})^{2}}{64\pi \bar{A}_{N}}. \label{normalized_luminosity}
\end{equation}

It is interesting to express the black hole luminosity in terms of
black hole classical mass~$M$
\begin{equation}
L \sim \delta\omega_{0}^{2} = \frac{1}{1024 \pi^{2}M^{2}}\gamma,
\end{equation}
where 
\begin{equation}
\gamma = \big(4\pi R^{2}\big)^{2}(\bar{A}_{N} - \bar{A}_{N-1})^{2}.
\end{equation}
As we can see, as the black hole shrinks the factor~$\gamma$ deviates
increasingly from a constant value, in a way that the black hole
luminosity also deviates increasingly from Page's~$M^{-2}$
semiclassical result. Thus, the factor~$\gamma$ becomes important as
the black hole becomes smaller in order to confirm the quantum nature
of the spacetime. 

\begin{figure}
\centering
\includegraphics[scale=0.5]{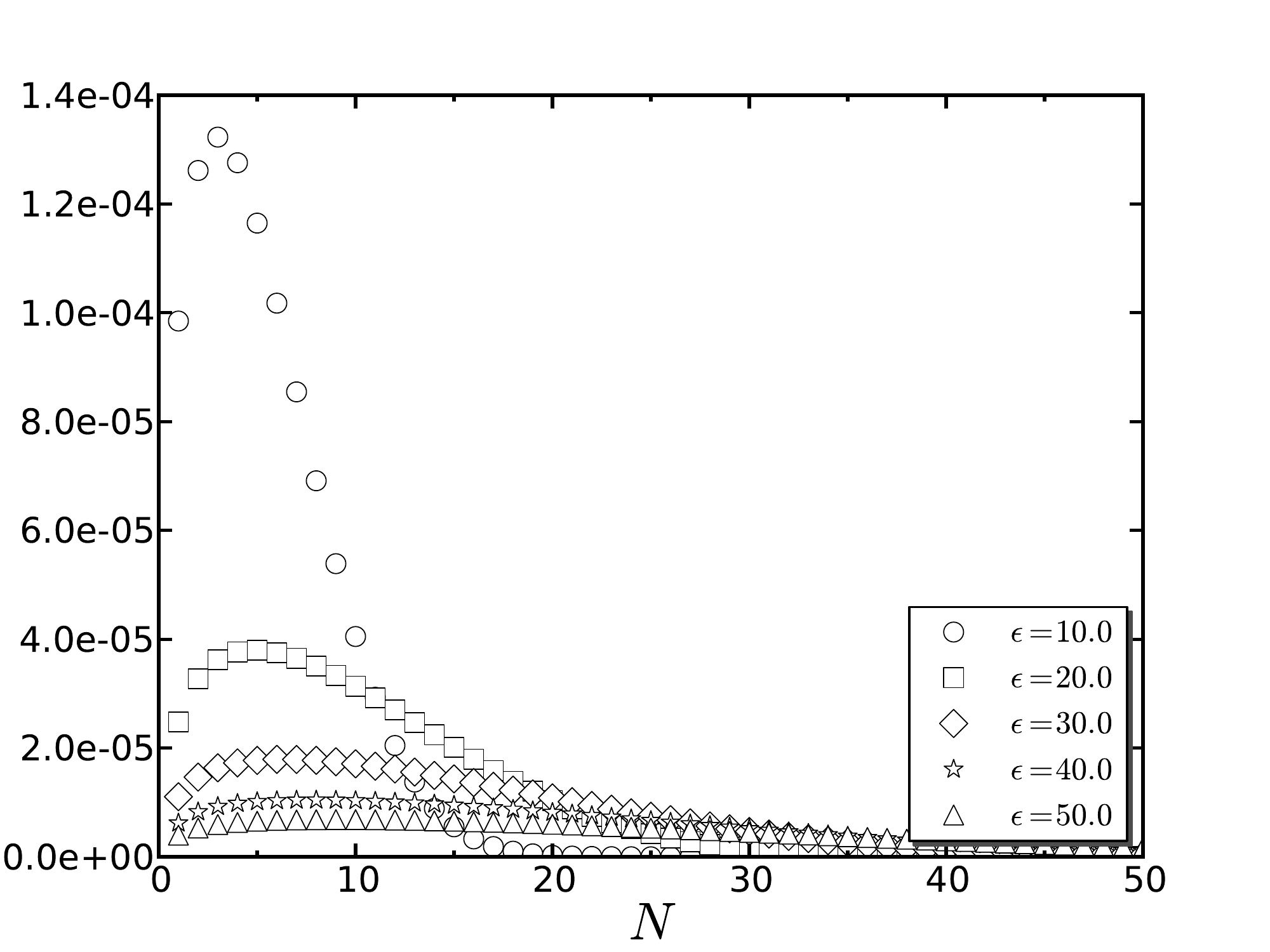}
\caption{Normalized luminosity~\eqref{normalized_luminosity} as
  a function of~$N$, for a fixed energy scale cut-off
  scale~$\epsilon$. Here again, notice the effect of the choice of the cut-off scale in the
  significance of the noncommutativity on the luminosity, where
  the quantum corrections contribute increasingly.}
\label{fig:norm_luminosity}
\end{figure}

Black hole normalized luminosity is shown in the figure ~\eqref{fig:norm_luminosity}. Since the luminosity increases as the fundamental
energy scale~$\epsilon$ increases, non-commutative corrections can
also be observed if the fundamental energy scale could be probed, like
in large extra-dimensional models where particle collision at LHC can
produce micro-black holes~\cite{2001-Dimopoulos-PRL-161602,
  2003-Cavaglia-CQG-205}.

\section{Conclusions}
In this paper we have performed an investigation on the area spectrum
of a 2-sphere, with a correction due to quantum fluctuations of space
through the asymptotic expansion for the heat kernel trace. This
investigation can shed some light on the nature of spacetime in the
Planck scale which can be reflected in quantum gravity corrections on
the luminosity of the black holes, and verified in the LHC.

We can verify that, in our approach, in order to observe a correction for the area of
order of~$1\%$ we should have $\frac{N(N+1)}{R^{2}}\sim 6.87\times 10^{56}\,\text{eV}^2$.
In this way, using the estimated Schwarschild radius~$R\sim
10^{20}\unit{eV}^{-1}$ for the M87 black
hole~\cite{2012-Doeleman-S-355} we conclude that~$N=10^{50}$ still
corresponds to a strong quantum regime. It indicates that ~$N$ can be interpreted 
as the number of quanta area~$N\sim
A/\ell\textsubscript{P}$ as occurs in entropic
gravity approach ~\cite{2013-Gregory-PRD-64030}.

The quantum gravity corrections found out become stronger as we get
close to the fundamental cutoff scale, and since such corrections
decrease the area, we should expect that our approach breaks close to
the Planck scale, where a full theory of quantum gravity will be
necessary. In despite of this, the results obtained in this article
can shed some light on the quantum nature of the spacetime in the
Planck scale.  Other ingredients, like extra dimensions, can be added
later in order to provide more precise experimental signatures of
noncommutativity.

\section{Acknowledgments}
The authors thank the Brazilian Agencies \emph{Conselho Nacional de
  Desenvolvimento Cient\'ifico e Tecnol\'ogico}~(CNPq) and
\emph{Coordena\c c\~ao de Aperfei\c coamento de Pessoal de N\'ivel
  Superior}~(CAPES) for the financial support.  Victor Santos would
like to thank Professor Abhay Ashtekar from the \emph{Institute of
  Gravitation and the Cosmos} of the Pennsylvania State University for
the hospitality during the preparation of the manuscript.

\bibliographystyle{elsarticle-num}
\bibliography{fuzzyarea}

\begin{thebibliography}{10}
\expandafter\ifx\csname url\endcsname\relax
  \def\url#1{\texttt{#1}}\fi
\expandafter\ifx\csname urlprefix\endcsname\relax\def\urlprefix{URL }\fi
\expandafter\ifx\csname href\endcsname\relax
  \def\href#1#2{#2} \def\path#1{#1}\fi

\bibitem{1985-Heisenberg--382}
W.~Heisenberg, Über quantentheoretische umdeutung kinematischer und
  mechanischer beziehungen, in: W.~Blum, H.~Rechenberg, H.-P. Dürr (Eds.),
  Original Scientific Papers Wissenschaftliche Originalarbeiten, Vol. A / 1 of
  Werner Heisenberg Gesammelte Werke Collected Works, Springer Berlin
  Heidelberg, 1985, pp. 382--396.

\bibitem{1979-Pauli--}
W.~Pauli, A.~Hermann, K.~Meyenn, V.~Weisskopf, Scientific correspondence with
  Bohr, Einstein, Heisenberg, a.o, Sources in the history of mathematics and
  physical sciences, Springer, 1979.

\bibitem{1947-Snyder-PR-38}
H.~S. Snyder, Quantized space-time, Phys. Rev. 71 (1947) 38--41.

\bibitem{1947-Snyder-PR-68}
H.~S. Snyder, The electromagnetic field in quantized space-time, Phys. Rev. 72
  (1947) 68--71.

\bibitem{1947-Yang-PR-874}
C.~N. Yang, On quantized space-time, Phys. Rev. 72 (1947) 874--874.

\bibitem{1995-Connes--}
A.~Connes, Noncommutative Geometry, Elsevier Science, 1995.

\bibitem{1999-Madore--}
J.~Madore, An Introduction to Noncommutative Differential Geometry and Its
  Physical Applications, London Mathematical Society Lecture Note Series,
  Cambridge University Press, 1999.

\bibitem{1998-Connes-JHEP-3}
A.~Connes, M.~Douglas, A.~Schwarz, Noncommutative geometry and matrix theory,
  JHEP 1998~(02) (1998) 003.

\bibitem{2000-Castellani-CQG-3377}
L.~Castellani, Non-commutative geometry and physics: a review of selected
  recent results, Class. Quantum Grav. 17~(17) (2000) 3377.

\bibitem{1999-Seiberg-JHEP-32}
N.~Seiberg, E.~Witten, String theory and noncommutative geometry, JHEP
  1999~(09) (1999) 032.

\bibitem{1998-Cheung-NPB-185}
Y.-K.~E. Cheung, M.~Krogh, Non-commutative geometry from 0-branes in a
  background b-field, Nucl. Phys. B 528~(1--2) (1998) 185--196.

\bibitem{1999-Chu-NPB-151}
C.-S. Chu, P.-M. Ho, Non-commutative open string and d-brane, Nucl. Phys. B
  550~(1--2) (1999) 151--168.

\bibitem{1999-Schomerus-JHEP-30}
V.~Schomerus, D-branes and deformation quantization, JHEP 1999~(06) (1999) 030.

\bibitem{2000-Seiberg-JHEP-21}
N.~Seiberg, L.~Susskind, N.~Toumbas, Strings in background electric field,
  space/time noncommutativity and a new noncritical string theory, JHEP
  2000~(06) (2000) 021.

\bibitem{2000-Seiberg-JHEP-44}
N.~Seiberg, L.~Susskind, N.~Toumbas, Space/time non-commutativity and
  causality, JHEP 2000~(06) (2000) 044.

\bibitem{2011-Baratin-CQG-175011}
A.~Baratin, B.~Dittrich, D.~Oriti, J.~Tambornino, Non-commutative flux
  representation for loop quantum gravity, Class. Quantum Grav. 28~(17) (2011)
  175011.

\bibitem{2011-Aastrup-CQG-75014}
J.~Aastrup, J.~M. Grimstrup, M.~Paschke, Quantum gravity coupled to matter via
  noncommutative geometry, Class. Quantum Grav. 28~(7) (2011) 075014.

\bibitem{2010-Denicola-CQG-205025}
D.~Denicola, M.~Marcolli, A.~Z. al~Yasry, Spin foams and noncommutative
  geometry, Class. Quantum Grav. 27~(20) (2010) 205025.

\bibitem{1929-Robertson-PR-163}
H.~P. Robertson, The uncertainty principle, Phys. Rev. 34 (1929) 163--164.

\bibitem{2006-Freidel-PRL-221301}
L.~Freidel, E.~R. Livine, 3d quantum gravity and effective noncommutative
  quantum field theory, Phys. Rev. Lett. 96 (2006) 221301.

\bibitem{Nicolini:2008aj}
P.~Nicolini, Noncommutative black holes, the final appeal to quantum gravity: A
  review, Int. J. Mod. Phys. A 24 (2009) 1229--1308.

\bibitem{2005-Dolan-JHEP-8}
B.~P. Dolan, Quantum black holes: the event horizon as a fuzzy sphere, JHEP
  2005~(02) (2005) 008.

\bibitem{2009-Silva-PLB-318}
C.~A.~S. Silva, Fuzzy spaces topology change as a possible solution to the
  black hole information loss paradox, Phys. Lett. B 677~(5) (2009) 318--321.

\bibitem{2011-Silva-EPL-10007}
C.~A.~S. Silva, R.~R. Landim, A note on black-hole entropy, area spectrum, and
  evaporation, Europhys. Lett. 96~(1) (2011) 10007.

\bibitem{Hod:1998vk}
S.~Hod, Bohr's correspondence principle and the area spectrum of quantum black
  holes, Phys. Rev. Lett. 81 (1998) 4293--4296.

\bibitem{Banerjee:2010be}
R.~Banerjee, B.~R. Majhi, E.~C. Vagenas, A note on the lower bound of
  black-hole area change in the tunneling formalism, Europhys. Lett. 92~(2)
  (2010) 20001.

\bibitem{Rovelli:1994ge}
C.~Rovelli, L.~Smolin, Discreteness of area and volume in quantum gravity,
  Nucl. Phys. B 442~(3) (1995) 593 -- 619.

\bibitem{1994-Gilkey--}
P.~B. Gilkey, Invariance Theory, the Heat Equation, and the Atiyah-Singer Index
  Theorem, Studies in Advanced Mathematics, CRC Press, 1994.

\bibitem{1967-Mckean-JoDG-43}
H.~P. McKean, I.~M. Singer, Curvature and the eigenvalues of the laplacian, J.
  Diff. Geom. 1~(1--2) (1967) 43--69.

\bibitem{1994-Connes--}
A.~Connes, Noncommutative Geometry, Academic Press, 1994.

\bibitem{1992-Madore-CQG-69}
J.~Madore, The fuzzy sphere, Class. Quantum Grav. 9~(1) (1992) 69.

\bibitem{2006-Sheikh-Jabbari-PLB-119}
M.~M. Sheikh-Jabbari, Inherent holography in fuzzy spaces and an n-tropic
  approach to the cosmological constant problem, Phys. Lett. B 642~(1-–2)
  (2006) 119--123.

\bibitem{2003-Abe-PRD-25002}
Y.~Abe, V.~P. Nair, Noncommutative gravity: Fuzzy sphere and others, Phys. Rev.
  D 68 (2003) 025002.

\bibitem{2003-OConnor-MPLA-2423}
D.~O'Connor, Field theory on low dimensional fuzzy spaces, Mod. Phys. Lett. A
  18~(33n35) (2003) 2423--2430.

\bibitem{2012-Doeleman-S-355}
S.~S. Doeleman, V.~L. Fish, D.~E. Schenck, C.~Beaudoin, R.~Blundell, G.~C.
  Bower, A.~E. Broderick, R.~Chamberlin, R.~Freund, P.~Friberg, M.~A. Gurwell,
  P.~T.~P. Ho, M.~Honma, M.~Inoue, T.~P. Krichbaum, J.~Lamb, A.~Loeb,
  C.~Lonsdale, D.~P. Marrone, J.~M. Moran, T.~Oyama, R.~Plambeck, R.~A.
  Primiani, A.~E.~E. Rogers, D.~L. Smythe, J.~SooHoo, P.~Strittmatter, R.~P.~J.
  Tilanus, M.~Titus, J.~Weintroub, M.~Wright, K.~H. Young, L.~M. Ziurys,
  {Jet-Launching Structure Resolved Near the Supermassive Black Hole in M87},
  Science 338~(6105) (2012) 355--358.

\bibitem{2013-Gregory-PRD-64030}
C.~Gregory, A.~Pinzul, {On Noncommutative Effects in Entropic Gravity}, Phys.
  Rev. D 88 (2013) 064030.

\bibitem{1976-Page-PRD-198}
D.~N. Page, Particle emission rates from a black hole: Massless particles from
  an uncharged, nonrotating hole, Phys. Rev. D 13 (1976) 198--206.

\bibitem{1976-Page-PRD-3260}
D.~N. Page, Particle emission rates from a black hole. ii. massless particles
  from a rotating hole, Phys. Rev. D 14 (1976) 3260--3273.

\bibitem{1977-Page-PRD-2402}
D.~N. Page, Particle emission rates from a black hole. iii. charged leptons
  from a nonrotating hole, Phys. Rev. D 16 (1977) 2402--2411.

\bibitem{1997-Makela-PLB-115}
J.~M\"akel\"a, Black hole spectrum: continuous or discrete?, Phys. Lett. B
  390~(1--4) (1997) 115--118.

\bibitem{2001-Dimopoulos-PRL-161602}
S.~Dimopoulos, G.~Landsberg, Black holes at the large hadron collider, Phys.
  Rev. Lett. 87 (2001) 161602.

\bibitem{2003-Cavaglia-CQG-205}
M.~Cavagli\`a, S.~Das, R.~Maartens, {Will we observe black holes at the LHC?},
  Class. Quantum Grav. 20~(15) (2003) L205.

\end{thebibliography}

\end{document}